\documentclass[journal]{IEEEtran}

\usepackage{cite}
\usepackage{threeparttable}
\usepackage{graphicx}
\usepackage{picinpar}
\usepackage[cmex10]{amsmath}
\usepackage{amsmath,amsfonts,amssymb}
\usepackage{subfigure}
\usepackage{enumerate}

\usepackage{cite}
\usepackage{threeparttable}
\usepackage{graphicx}
\usepackage{picinpar}
\usepackage[cmex10]{amsmath}
\usepackage{amsmath,amsfonts,amssymb}
\usepackage{subfigure}
\usepackage{algorithm}
\usepackage{algorithmic}
\usepackage{stfloats}
\usepackage{bm}
\usepackage{amsthm}
\usepackage{changebar}
\usepackage{algorithm}
\usepackage{algorithmic}
\usepackage{stfloats}
\usepackage{bm}
\usepackage{mathrsfs}
\usepackage{color}

\begin{document}

\bibliographystyle{IEEEtran}

\title{MmWave Massive MIMO Based Wireless Backhaul for 5G Ultra-Dense Network}
\author{Zhen Gao, Linglong Dai, De Mi, Zhaocheng Wang, Muhammad Ali Imran, and Muhammad Zeeshan Shakir
%
\thanks{Z. Gao, L. Dai, and Z. Wang are with Tsinghua National Laboratory for
 Information Science and Technology (TNList), Department of Electronic Engineering,
 Tsinghua University, Beijing 100084, P. R. China (E-mails: gao-z11@mails.tsinghua.edu.cn;
\{daill, zcwang\}@tsinghua.edu.cn).}
\thanks{D. Mi and M. A. Imran are with Institute for Communication Systems (ICS), Home of 5G Innovation Center (5GIC), University of Surrey, Guildford, UK (E-mails: \{d.mi, m.imran\}@surrey.ac.uk).} %
\thanks{M. Z. Shakir is with the Department of Electrical and Computer Engineering, Texas A\&M University at Qatar (TAMUQ), Education City, P.O. Box
23874, Doha, Qatar (E-mail: muhammad.shakir@qatar.tamu.edu).}
\thanks{This work was supported in part by the International Science \& Technology Cooperation Program of China (Grant No. 2015DFG12760),
the National Natural Science Foundation of China (Grant No. 61201185 and 61271266),  the Beijing Natural Science Foundation (Grant No. 4142027),
and the Foundation of Shenzhen government.}

\vspace*{-5.0mm}
}

\maketitle

\begin{abstract}
Ultra-dense network (UDN) has been considered as a promising candidate for future 5G network to meet the explosive data demand. To realize UDN, a reliable, Gigahertz bandwidth, and cost-effective backhaul connecting ultra-dense small-cell base stations (BSs) and macro-cell BS is prerequisite. Millimeter-wave (mmWave) can provide the potential Gbps traffic for wireless backhaul. Moreover, mmWave can be easily integrated with massive MIMO for the improved link reliability. In this article, we discuss the feasibility of mmWave massive MIMO based wireless backhaul for 5G UDN, and the benefits and challenges are also addressed. Especially, we propose a digitally-controlled phase-shifter network (DPSN) based hybrid precoding/combining scheme for mmWave massive MIMO, whereby the low-rank property of mmWave massive MIMO channel matrix is leveraged to
reduce the required cost and complexity of transceiver with a negligible performance loss. One key feature of the proposed scheme is that the macro-cell BS can simultaneously support multiple small-cell BSs with multiple streams for each small-cell BS, which is essentially different from conventional hybrid precoding/combining schemes typically limited to single-user MIMO with multiple streams or multi-user MIMO with single stream for each user.
Based on the proposed scheme, we further explore the fundamental issues of developing mmWave massive MIMO for wireless backhaul, and the associated challenges, insight, and prospect to enable the mmWave massive MIMO based wireless backhaul for 5G UDN are discussed.
\end{abstract}


\vspace*{-3.0mm}
\section{Introduction}\label{S1}
\IEEEPARstart{T}{he} explosive traffic demand is challenging current cellular networks, including the most advanced 4G network. It has been the consensus that future 5G network should realize the goals of thousand-fold system capacity, hundred-fold energy efficiency, and tens of lower latency. To realize such aggressive 5G version, ultra-dense network (UDN) has been considered as a promising system architecture to enable Gbps user experience, seamless coverage, and green communications~\cite{5G}.

\begin{figure}[!tb]
     \centering
     \includegraphics[width=7.3cm]
     {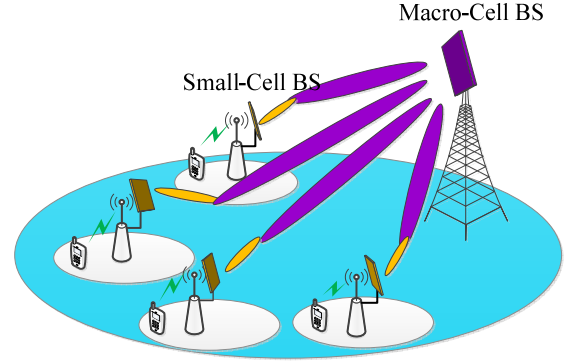}
     \vspace*{-2.0mm}
    \caption{MmWave massive MIMO based wireless backhaul for 5G UDN.}     \label{udn}
    \vspace*{-4mm}
\end{figure}

In UDN, as shown in Fig. \ref{udn}, the macro-cell base stations (BSs) with large coverage usually control the user scheduling, resource allocation, and support high-mobility users, while many ultra-dense small-cell BSs with much smaller coverage provide the high data rate for low-mobility users. Due to ultra-dense small-cell BSs, better frequency reuse can be achieved, and energy efficiency can be also substantially improved due to the reduced path loss in small cells~\cite{5G}.

To enable UDN, a reliable, cost-effective, and Gigahertz bandwidth backhaul connecting macro-cell BS and the associated small-cell BSs is prerequisite. It has been demonstrated that backhaul with 1$\sim$10 GHz bandwidth is required to effectively support UDN \cite{in_band}. Conventional optical fiber enjoys large bandwidth and reliability, but its application to UDN as backhaul may not be an economical choice for operators due to the restriction of deployment and installation. Hence, wireless backhaul, especially millimeter-wave (mmWave) backhaul, is more attractive to overcome the geographical constraints. The advantages of mmWave backhaul are:
\begin{itemize}
  \item \emph{A large amount of underutilized band }in mmWave can be leveraged to provide the potential Gigahertz transmission bandwidth, which is different from scarce microwave band in conventional cellular networks~\cite{key}.
  \item \emph{A large number of antennas} can be easily employed for mmWave communications due to the small wavelength of mmWave, which can improve the signal directivity (reduce the co-channel interference) and link reliability (mitigate the large path loss) for mmWave backhaul~\cite{Y_hybrid}.
\end{itemize}

This article combines mmWave with a large number of antennas, which is also referred to as mmWave massive MIMO, to provide wireless backhaul for future 5G UDN. The contributions of this article are listed as follows:
\begin{itemize}
  \item We discuss the feasibility and challenges of the mmWave massive MIMO based backhaul for UDN, where its advantages, differences compared with conventional massive MIMO working at sub 3$\sim$6 GHz for radio access networks (RAN) are also addressed. Moreover, the sparsity of mmWave massive MIMO channels is stressed.
  \item We explore key issues and potential research directions of the cost-effective mmWave massive MIMO for UDN backhaul. Especially, a digitally-controlled phase-shifter network (DPSN) based hybrid precoding/combining and the associated compressive sensing (CS) based channel estimation is proposed.

  \item We address the benefits of the wireless backhaul for 5G UDN with the technique of mmWave massive MIMO, which may provide a viable approach to realize the novel backhaul network topology, scheduling strategy, efficient in-band backhaul in mmWave.
\end{itemize}
\vspace*{-4mm}
\section{Feasibility and Challenges of MmWave Massive MIMO for Wireless Backhaul in 5G UDN}%

In UDN, small cells are densely deployed in hotspots (e.g., office buildings, shopping malls, resident apartments) with high data rate to provide traffic offload from macro cells, since the large majority of traffic demand comes from these hotspots. Hence, the backhaul between the macro-cell BS and the associated small-cell BSs should provide large bandwidth with reliable link transmission. Besides, power efficiency and deployment cost are also key considerations for operators.
\vspace*{-3mm}
\subsection{MmWave is Suitable for Wireless Backhaul in 5G UDN}%
Traditionally, mmWave is not used for RAN in existing cellular networks due to its high path loss and expensive electron components. However, mmWave is especially suitable for backhaul in UDN due to the following reasons.
\begin{itemize}
  \item \emph{High Capacity and Inexpensive}: The large amount of underutilized mmWave including unlicensed V-band (57-67GHz) and lightly       licensed E-band (71-76GHz and 81-86GHz) (the specific regulation may vary from country to country) can provide the potential Gigahertz transmission bandwidth \cite{key}. For example, more than one Gbps backhaul capacity can be supported over 250 MHz channel in E-band \cite{in_band}.
  \item \emph{Immunity to Interference}: Transmission distance comfort zone for E-band is up to several kilometers due to the rain attenuation, while that for V-band is about 500-700m due to both the rain and oxygen attenuation. Owing to the high path loss, mmWave is suitable for UDN, where the improved frequency reuse and reduced inter-cell interference are expected. It should be pointed out that rain attenuation is not a big issue for mmWave used in UDN. If we consider the very heavy rainfall of 25mm/hr, the rain attenuation is only around 2 dB in E-band if we consider the distance of backhaul link is 200m in typical urban UDN \cite{key}.
  \item \emph{Small Form Factor}: The small wavelength of mmWave implies that massive antennas can be easily equipped at both macro and small-cell BSs, which can improve the signal directivity and compensate severe path loss of mmWave to achieve larger coverage in turn \cite{Y_hybrid}. Hence the compact mmWave backhaul equipment can be easily deployed with low cost sites (such as light poles, building walls, bus stations) and short installation time.
\end{itemize}

\vspace*{-3mm}
\subsection{MmWave Massive MIMO is Different from Microwave Massie MIMO} \label{different}
Inheriting the advantages from conventional microwave massive MIMO, mmWave massive MIMO has the flexible beamforming, spatial multiplexing, and diversity. Hence mmWave massive MIMO brings not only the improved reliability of backhaul link, but also new architecture of backhaul network including the flexible network topology, scheduling scheme, which will be further detailed in Section \ref{benefits}. However, compared with conventional microwave (sub 3$\sim$6 GHz) massive MIMO used for RAN, the implementation of mmWave massive MIMO also brings challenges as follows.
\begin{itemize}
  \item
  First, the cost and complexity of transceiver including high-speed analog-digital converters (ADCs) and digital-analog converters (DACs), synthesizers, mixers, etc., in mmWave communications are much larger than that in conventional microwave communications. Hence, massive low-cost antennas but a limited number of expensive baseband (BB) chains can be an appealing transceiver structure for mmWave massive MIMO, which, however, challenges conventional precoding/combining schemes.

  \item

  Second, the number of antennas in mmWave at both macro and small-cell BSs can be much larger than that in conventional microwave massive MIMO due to the much smaller wavelength of mmWave. This implies the challenge that channel estimation in mmWave massive MIMO can be more difficult even when time division duplex (TDD) leveraging the channel reciprocity is considered. Even for TDD-based mmWave communications, the synchronization and calibration error of radio frequency (RF) chains to guarantee the channel reciprocity are not trivial \cite{TDD}.

  \item
Third, since single-antenna users are typically considered in microwave massive MIMO due to the limited form factor,
only channel state information at transmitter (CSIT) is required for precoding. However, for mmWave massive MIMO
where each small-cell BS can be equipped with massive antennas, precoding in the uplink and combining in
the downlink at small-cell BSs are also necessary, since precoding/combining can effectively support multiple
streams and directional transmission for the improved link reliability. Therefore,
channel state information at receiver (CSIR) is also required for mmWave massive MIMO, which indicates another
challenge that channel estimation acquired in the uplink by leveraging the channel reciprocity should also be
feedback to small-cell BSs.
\end{itemize}

\vspace*{-3mm}
\section{MmWave Channel Characteristics}\label{No}

As discussed above, the mmWave massive MIMO based backhaul is apt to the transceiver with the limited number of BB chains. 
Compared with microwave massive MIMO using full digital precoding,
precoding/combining with the smaller number of BB chains than that of antenna elements can make mmWave massive MIMO suffer from a certain performance loss, which is largely dependent on the propagation condition of mmWave massive MIMO channels.

\setcounter{figure}{2}
\begin{figure*}[!b]
     \centering
     \vspace*{-0.5mm}
     \includegraphics[width=18cm]
     {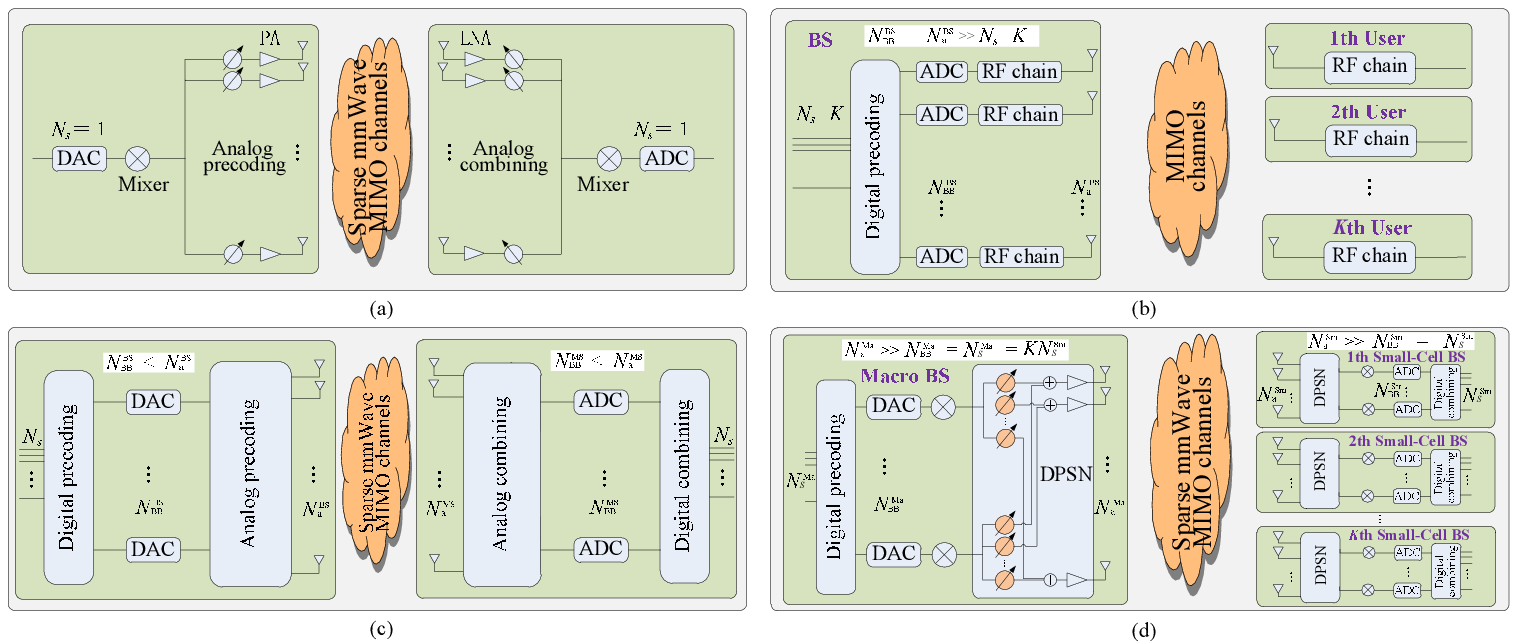}
     \vspace*{-2.0mm}
    \caption{Comparison of precoding/combining schemes, where PA denotes power amplifier and LNA denotes low-noise amplifier: (a) Analog precoding/combining scheme in mmWave multi-antenna systems; (b) Digital precoding in microwave massive MIMO for RAN; (c) Conventional hybrid precoding/combining in mmWave massive MIMO for RAN; (d) Proposed DPSN based hybrid precoding/combining for mmWave massive MIMO in UDN backhaul.}     \label{table}
    \vspace*{-3mm}
\end{figure*}

\setcounter{figure}{1}
\begin{figure}[!tb]
     \centering
     \includegraphics[width=9cm]
     {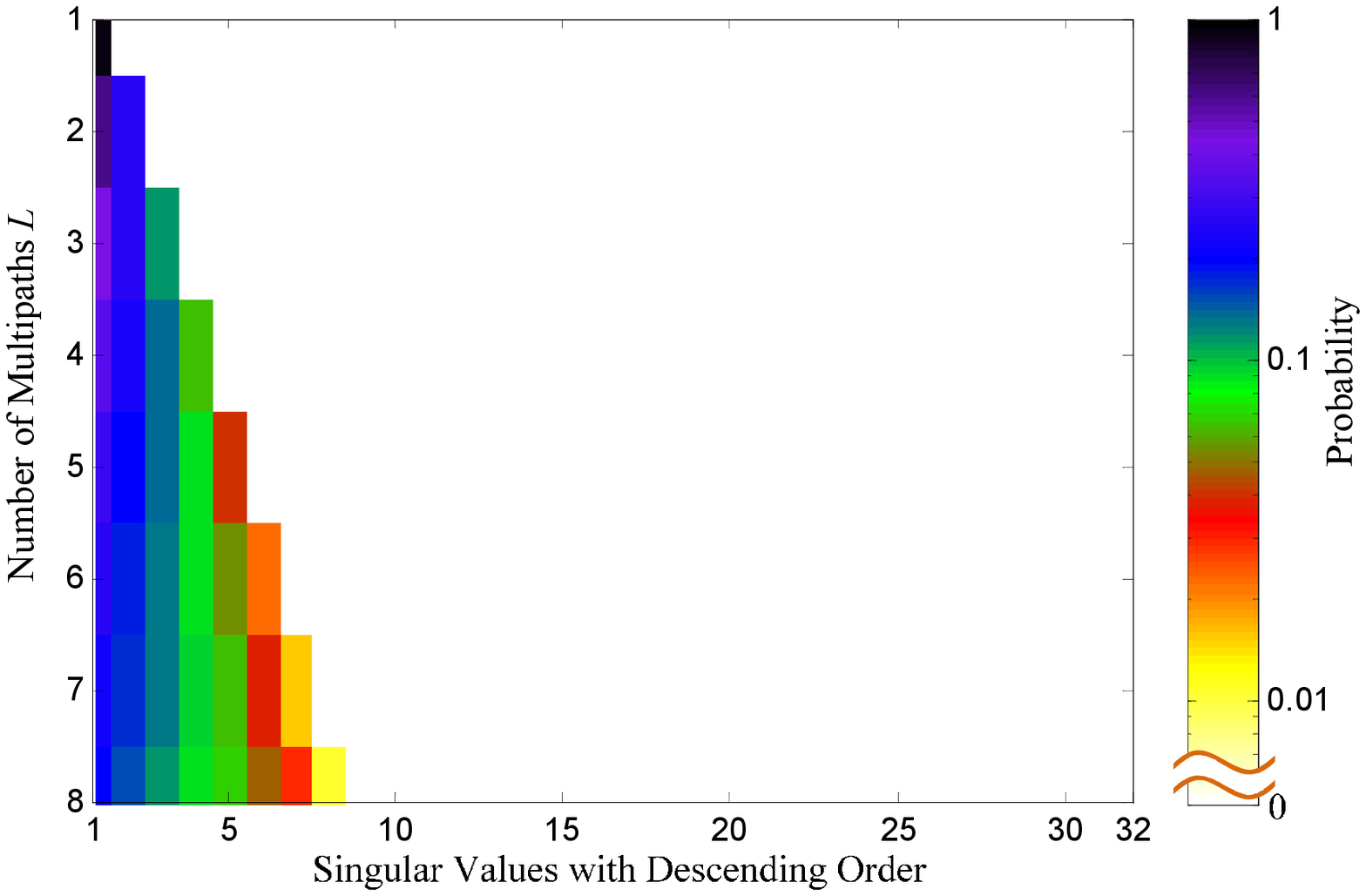}
    \vspace*{-5.0mm}
   \caption{Energy probability distribution of singular values with descending order of mmWave massive MIMO channel matrix versus different $L$'s, where $N_T=512$ and $N_R=32$.}     \label{low_rank}
   \vspace*{-6mm}
\end{figure}
\setcounter{figure}{3}
\subsection{MmWave Channels with Spatial/Angular Sparsity}
Extensive experiments have shown that mmWave massive MIMO channels exhibit the obviously spatial/angular sparsity due to its high path loss for non-line-of-sight (NLOS) signals, where only a small number of dominated multipaths (typically, 3$ \sim $5 multipaths in realistic environments~\cite{scanning}) consist of mmWave MIMO multipath channels. If we consider the widely used uniform linear array (ULA), the point-to-point mmWave massive MIMO channel can be modeled as~\cite{scanning}
\begin{equation}
{\bf{H}}{\rm{ = }}\sqrt {\frac{{{N_T}{N_R}}}{\rho }} \sum\limits_{l = 1}^L {{\alpha _l}{{\bf{a}}_{{T}}}\left( {{\theta _l}} \right){\bf{b}}_R^*\left( {{\varphi _l}} \right)} {\rm{ = }}\sqrt {\frac{{{N_T}{N_R}}}{\rho }} {\bf{A}}_T{\bf{D }}{{\bf{B}}_R^{\rm{*}}},\label{equ1}
\end{equation}
where $N_T$ and $N_R$ are the numbers of transmit and receive antennas, respectively, $\rho $ is the average path loss, $L$ is the number of multipaths, ${{\alpha _l}}$ is the complex gain of the $l$th path, ${{\theta _l}}\in [0,2\pi]$ and ${{\varphi _l}}\in [0,2\pi]$ are azimuth angles of departure or arrival (AoD/AoA). 
In addition, ${{\bf{a}}_T}\left( {{\theta _l}} \right) = \frac{1}{{\sqrt {{N_T}} }}{\left[ {1,{e^{j2\pi d\sin ({\theta _l})/\lambda }}, \cdots ,{e^{j2\pi ({N_T} - 1)d\sin ({\theta _l})/\lambda }}} \right]^{\rm{T}}}$ and ${{\bf{b}}_R}\left( {{\varphi  _l}} \right) = \frac{1}{{\sqrt {{N_R}} }}{\left[ {1,{e^{j2\pi d\sin ({\varphi  _l})/\lambda }}, \cdots ,{e^{j2\pi ({N_R} - 1)d\sin ({\varphi  _l}) /\lambda }}} \right]^{\rm{T}}}$ are steering vectors at the transmitter and receiver, respectively, ${{\bf{A}}_T} = \left[ {{{\bf{a}}_{{T}}}\left( {{\theta _1}} \right){\rm{|}}{{\bf{a}}_{{T}}}\left( {{\theta _2}} \right){\rm{|}} \cdots {\rm{|}}{{\bf{a}}_{{T}}}\left( {{\theta _L}} \right)} \right]$, ${{\bf{B}}_R}{\rm{ = }}\left[ {{{\bf{b}}_R}\left( {{\varphi _1}} \right){\rm{|}}{{\bf{b}}_R}\left( {{\varphi _2}} \right){\rm{|}} \cdots {\rm{|}}{{\bf{b}}_R}\left( {{\varphi _L}} \right)} \right]^{\rm{*}}$, and the diagonal matrix ${\bf{D }}={\rm{diag}}\left\{ {{\alpha _1},{\alpha _2}, \cdots ,{\alpha _L}} \right\}$, where $\lambda $ and $d$ are wavelength and antenna spacing, respectively.

\vspace*{-3.0mm}
\subsection{Low-Rank Property of MmWave Massive MIMO Channels}
The spatial/angular sparsity of mmWave channels with small $L$ (e.g., 3$ \sim $5) and massive MIMO channel matrix with large $N_T$, $N_R$ (dozens even hundreds) implies that mmWave massive MIMO channel matrix has the low-rank property~\cite{magzine_solution}. For example, Fig. 2 provides the energy probability distribution of singular values of $\bf{H}$ with descending order against different $L$'s, where $N_T=512$, $N_R=32$, and path gains follow the independent and identically distributed (i.i.d.) complex Gaussian distribution. It can be observed that the mmWave massive MIMO channel matrix has the obvious low-rank property. If we consider single user (SU)-MIMO with CSIT for precoding and CSIR for combining, the low-rank channel matrix indicates that the number of effective independent streams which can be exploited is small. Theoretical analysis has shown that the capacity of MIMO systems over sparse mmWave channels appears ceiling effect with the increased number of BB chains \cite{magzine_solution}. Hence, we can leverage the finite number of BB chains to maximize the backhaul capacity over sparse mmWave channels, where the number of BB chains can be as small as the effective rank of mmWave massive MIMO channel matrix.

\section{Key Issues of Designing MmWave Massive MIMO for 5G UDN Backhaul}\label{design}
\subsection{Hybrid Precoding/Combining Design}\label{precoding}
In order to realize the reliable point-to-multiple-points (P2MP) backhaul link, mmWave massive MIMO for UDN backhaul should exploit the flexible beamforming and spatial multiplexing to simultaneously support multiple small-cell BSs and provide multiple streams for each small-cell BS, which challenges conventional precoding/combining schemes.
\subsubsection{Overview of Existing Precoding/Combining Schemes}
Conventional mmWave multi-antenna systems utilize single RF chain and analog (e.g., ferrite based) phase-shifters for precoding/combining as shown in Fig. 3 (a), but it is limited to SU-MIMO with single stream. Full digital precoding in microwave massive MIMO, as shown in Fig. 3 (b), can simultaneously support multiple single-antenna users, i.e., MU-MIMO, but it requires one specific RF chain to be connected to each antenna, which can be unaffordable in mmWave communications~\cite{magzine_solution}. Recently, the hybrid precoding/combining scheme consisting of analog and digital precoding/combining as shown in Fig. 3 (c), has been proposed for mmWave massive MIMO with the reduced cost and complexity of transceiver. However, state-off-the-art hybrid precoding/combining schemes are usually limited to SU-MIMO with multiple streams or multi-user (MU)-MIMO with single stream for each user~\cite{magzine_solution,scanning,Y_hybrid,Y_hybrid_training}.

\subsubsection{Proposed DPSN Based Hybrid Precoding/Combining: Multi-User and Multi-Stream}
To support multi-user and multi-stream, we propose the DPSN based hybrid precoding/combining scheme as shown in Fig. 3~(d), which can effectively reduce the cost and complexity of transceiver. Specifically, consider the macro-cell BS has $N_{\rm{a}}^{\rm{Ma}}$ antennas but $N_{\rm{BB}}^{\rm{Ma}}$ BB chains, where $N_{\rm{a}}^{\rm{Ma}} \gg N_{\rm{BB}}^{\rm{Ma}}$, while each small-cell BS has $N_{\rm{a}}^{\rm{Sm}}$ antennas but $N_{\rm{BB}}^{\rm{Sm}}$ BB chains, where $N_{\rm{a}}^{\rm{Sm}} \gg N_{\rm{BB}}^{\rm{Sm}}$. The number of simultaneously supported small-cell BSs is $K$. ${\bf{H}}_k \in {\mathbb{C}}^{N_{\rm{a}}^{\rm{Ma}} \times N_{\rm{a}}^{\rm{Sm}}}$ with $N_{\rm{a}}^{\rm{Ma}}>N_{\rm{a}}^{\rm{Sm}}$ denotes the mmWave massive MIMO channel matrix associated with the macro-cell BS and the $k$th small-cell BS, and it can be expressed as follows according to singular value decomposition (SVD):
\begin{equation}
{{\bf{H}}_k}{\rm{ = }}\left[ {{\bf{U}}_k^1{\rm{|}}{\bf{U}}_k^2} \right]\left[ {\begin{array}{*{20}{c}}
{{\bf{\Sigma }}_k^1}&{\bf{0}}\\
{\bf{0}}&{{\bf{\Sigma }}_k^2}\\
{\bf{0}}&{\bf{0}}
\end{array}} \right]\left[ {\begin{array}{*{20}{l}}
{{{\left( {{\bf{V}}_k^1} \right)}^{\rm{*}}}}\\
{{{\left( {{\bf{V}}_k^2} \right)}^{\rm{*}}}}
\end{array}} \right] \approx {\bf{U}}_k^1{\bf{\Sigma }}_k^1{\left( {{\bf{V}}_k^1} \right)^{\rm{*}}},\label{equ2}
\end{equation}
where both $\left[ {{\bf{U}}_k^1{\rm{|}}{\bf{U}}_k^2} \right]\in {\mathbb{C}}^{N_{\rm{a}}^{\rm{Ma}} \times N_{\rm{a}}^{\rm{Ma}}}$ and ${\left[ {{\bf{V}}_k^1|{\bf{V}}_k^2} \right]^{\rm{*}}}\in {\mathbb{C}}^{N_{\rm{a}}^{\rm{Sm}} \times N_{\rm{a}}^{\rm{Sm}}}$ are unitary matrices, ${{\bf{\Sigma }}_k^1}\in {\mathbb{C}}^{R_k \times R_k}$ and ${\bf{\Sigma }}_k^2 \in {{\mathbb{C}}^{\left( {N_{\rm{a}}^{{\rm{Sm}}} - {R_k}} \right) \times \left( {N_{\rm{a}}^{{\rm{Sm}}} - {R_k}} \right)}}$ are diagonal matrices whose diagonal elements are singular values of ${{\bf{H}}_k}$, and $R_k$ is the effective rank of ${\bf{H}}_k$. The approximation in (\ref{equ2}) is due to the low-rank property of ${\bf{H}}_k$ with ${\bf{\Sigma }}_k^2 \approx {\bf{0}}$, so that ${{\bf{U}}_k^1}\in {\mathbb{C}}^{N_{\rm{a}}^{\rm{Ma}} \times R_k}$ and ${\left( {{\bf{V}}_k^1} \right)^{\rm{*}}}\in {\mathbb{C}}^{R_k \times N_{\rm{a}}^{\rm{Sm}}}$.

Eq. (\ref{equ2}) indicates that $N_{\rm{BB}}^{\rm{Ma}}$ and $N_{\rm{BB}}^{\rm{Sm}}$ can be reduced to $R_k$ in SU-MIMO due to only $N_s=R_k$ effective independent streams. 
Moreover, we can use the precoding matrix ${\bf{P}}_k=\left({\bf{U}}_k^1\right)^{\rm{*}}$ and the combining matrix ${\bf{C}}_k= {\bf{V}}_k^1$ to effectively realize the independent multi-stream transmission~\cite{TSE}. To achieve this goal, we can use the emerging low-cost silicon-based SiGe and CMOS based programmable DPSN \cite{coms} to realize partial precoding/combining in the analog RF. With the cascade of the digital precoding matrix ${\bf{P}}_{d,k}\in {\mathbb{C}}^{R_k \times R_k}$ (or combining matrix ${\bf{C}}_{d,k}\in {\mathbb{C}}^{R_k \times R_k}$) and analog precoding matrix ${\bf{P}}_{a,k}\in {\mathbb{C}}^{R_k  \times N_{\rm{a}}^{\rm{Ma}}}$ (or combining matrix ${\bf{C}}_{a,k}\in {\mathbb{C}}^{N_{\rm{a}}^{\rm{Sm}} \times  R_k  }$), we can use ${\bf{P}}_{d,k}{\bf{P}}_{a,k}$ (or ${\bf{C}}_{a,k}{\bf{C}}_{d,k}$) to approximate ${\bf{P}}_k$ (or ${\bf{C}}_k$). Consider the precoding for instance, we can use the following iterative approach to acquire ${{{\bf{P}}_{d,k}}}$ and ${{{\bf{P}}_{a,k}}}$ that can minimize $ {\left\| {{{\bf{P}}_{k}} - {{\bf{P}}_{d,k}}{{\bf{P}}_{a,k}}} \right\|_F}$ with the constraint that elements in ${{\bf{P}}_{a,k}}$ are constant modulus. We initialize that ${{{\bf{\tilde P}}}_k} \leftarrow {{\bf{P}}_k}$. Then, we perform the following operations iteratively until ${{{\bf{P}}_{a,k}}}$ and ${{{\bf{P}}_{d,k}}}$ converge: 1) every element of ${{{\bf{P}}_{a,k}}}$ has the same phase with the corresponding element in ${{{\bf{\tilde P}}}_k}$; 2) ${{\bf{P}}_{d,k}} \leftarrow {{\bf{P}}_k}{\left( {{{\bf{P}}_{a,k}}} \right)^\dag }$, 3) ${{{\bf{\tilde P}}}_{k}} \leftarrow {\left( {{{\bf{P}}_{d,k}}} \right)^\dag }{{\bf{P}}_k}$. Note that ${{{\bf{P}}_{a,k}}}$ always meets the constraint of constant modulus and ${\left( {} \right)^\dag }$ is the Moore-Penrose pseudoinverse. Similarly, we can acquire ${{{\bf{C}}_{d,k}}}$ and ${{{\bf{C}}_{a,k}}}$ according to ${{{\bf{C}}_{k}}}$ with the same approach.
Besides, some power allocation strategies such as waterfilling can be integrated in the digital baseband precoding/combining to further improve the achievable capacity. 

Furthermore, consider the downlink MU-MIMO, where the channel matrix between macro-cell BS and $K$ small-cell BSs can be denoted as ${\bf{H}}\in {\mathbb{C}}^{N_{\rm{a}}^{\rm{Ma}} \times KN_{\rm{a}}^{\rm{Sm}}}$, and it can be represented~as ${\bf{H}}{\rm{ = }}\left[ {{{\bf{H}}_1}{\rm{|}}{{\bf{H}}_2}{\rm{|}} \cdots {\rm{|}}{{\bf{H}}_K}} \right]$ with ${{\bf{H}}_k} \approx {\bf{U}}_k^1{\bf{\Sigma }}_k^1{\left( {{\bf{V}}_k^1} \right)^{\rm{*}}}$ for $1\le k \le K$ according to (\ref{equ2}). Hence we further obtain
\begin{equation}
\begin{array}{l}
{\bf{H}}  \approx \left[ {{\bf{U}}_1^1{\rm{|}}{\bf{U}}_2^1| \cdots {\rm{|}}{\bf{U}}_K^1} \right]\times {\rm{diag}} \left\{ {{\bf{\Sigma }}_1^1,{\bf{\Sigma }}_2^1, \cdots ,{\bf{\Sigma }}_K^1} \right\}\\
 {\kern 15pt}  \times {\rm{diag}}\left\{ {{{\left( {{\bf{V}}_1^1} \right)}^{\rm{*}}},{{\left( {{\bf{V}}_2^1} \right)}^{\rm{*}}}, \cdots ,{{\left( {{\bf{V}}_K^1} \right)}^{\rm{*}}}} \right\},
\end{array}\label{equ3}
\end{equation}
where ${\bf{H}}_k$ for $1\le k \le K$ are assumed to share the same effective rank $R_k=R$. For precoding/combining in the proposed MU-MIMO system, the analog precoding matrix at macro-cell BS is ${{\bf{P}}_a} = {\left[ {{{\bf{P}}_{a,1}^T}{\rm{|}}{{\bf{P}}_{a,2}^T}| \cdots {\rm{|}}{{\bf{P}}_{a,K}^T}} \right]^T} \in { {\mathbb{C}}^{KR \times N_{\rm{a}}^{{\rm{Ma}}}}}$, and the analog and digital combining matrices for the $k$th small-cell BS can be ${\bf{C}}_{a,k}$ and ${\bf{C}}_{d,k}$, respectively. 
To further eliminate the multi-user interference, digital precoding ${{\bf{P}}_d} = {\left( {{{{\bf{\tilde P}}}_d}{{\bf{P}}_a}{\bf{\tilde U}}} \right)^{ - 1}}$ is proposed at the macro-cell BS, where ${{{\bf{\tilde P}}}_d} = {\rm{diag}}\left\{ {{{\bf{P}}_{d,1}},{{\bf{P}}_{d,2}}, \cdots ,{{\bf{P}}_{d,K}}} \right\}$ and ${\bf{\tilde U}} = \left[ {{\bf{U}}_1^1{\rm{|}}{\bf{U}}_2^1| \cdots {\rm{|}}{\bf{U}}_K^1} \right]$. The precoding/combining in the uplink of mmWave massive MIMO based backhaul is similar to the downlink, which will not be detailed in this article owing to the space limitation. 

The proposed precoding/combining scheme can diagonalize the equivalent channel ${{\bf{P}}_d}{{\bf{P}}_a}{\bf{H}}{\rm{diag}}\left\{ {{{\bf{C}}_{1}},{{\bf{C}}_{2}}, \cdots ,{{\bf{C}}_{K}}} \right\}$ with ${{\bf{C}}_{k}}={{\bf{C}}_{a,k}}{{\bf{C}}_{d,k}}$ to realize multi-user and multi-stream transmission, which is essentially different from existing schemes. 
Moreover, thanks to the obvious low-rank property of mmWave massive MIMO channel matrix as shown in Fig. 2, the proposed precoding/combining with the reduced number of BB chains only suffers from a negligible performance loss, which will be shown in Section \ref{Performance_Comparison}.

\vspace*{-3.0mm}
\subsection{CSI Acquisition for MmWave Massive MIMO}\label{CE}
To effectively realize the proposed DPSN based hybrid precoding/combining scheme, a reliable CSI acquisition scheme with low overhead is another challenge.
\begin{figure*}[!t]
     \centering
     \vspace*{-3.0mm}
     \includegraphics[width=17cm]
     {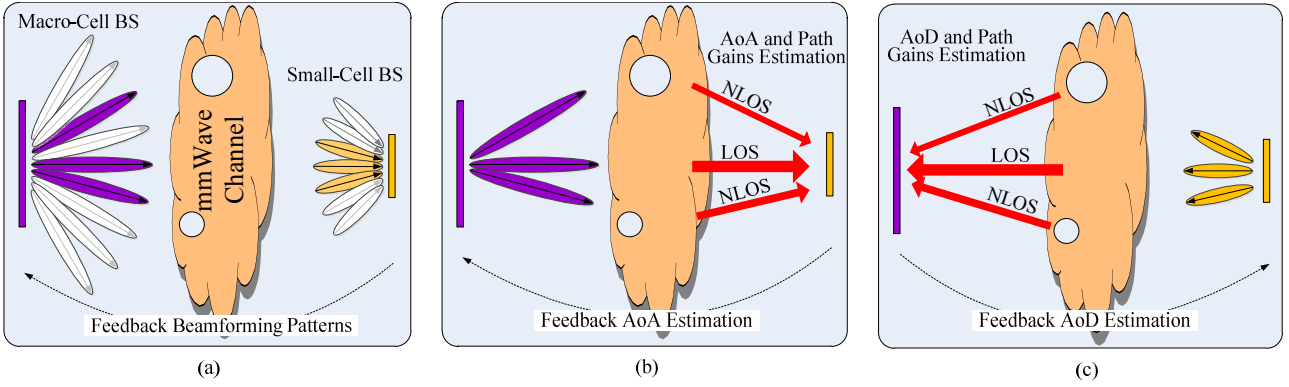}
     \vspace*{-3.0mm}
    \caption{CS-based channel estimation for mmWave massive MIMO based UDN backhaul: (a) Coarse channel estimation; (b) AoA and path gains estimation at small-cell BS; (c) AoD and path gains estimation at macro-cell BS.}     \label{fig:CE}
    \vspace*{-5mm}
\end{figure*}

\subsubsection{Challenging Channel Estimation for MmWave Massive MIMO}
As we have discussed in Section \ref{different}, mmWave massive MIMO may suffers from the prohibitively high overhead for channel estimation, and calibration error of RF chains as well as synchronization are also not trivial in TDD. Additionally, due to the much smaller number of BB chains than that of antennas, the effective dimensions that can be exploited for channel estimation will be substantially reduced although massive antennas are employed. Furthermore, channel estimation in the digital baseband should consider the characteristics of phase-shifter networks at both macro-cell BS and small-cell BSs, which can make it more complex. Finally, due to the strong signal directivity of mmWave, reliable channel estimation requires the sufficient received signal power, which means at least partial CSIT is necessary to ensure beamforming at the transmitter to match mmWave MIMO channels.

\subsubsection{Overview of Existing Channel Estimation Schemes}
Wireless local area networks (WLANs) (IEEE 802. 11ad) relies on the beamforming training to compensate the large path loss in 60 GHz \cite{magzine_solution}. The specific training consists of three phases: 1) sector level sweep is to select the best transmit and optionally receive antenna sector; 2) beam refinement is used for fine adjustment of beamforming; 3) beam tracking can adjust beamforming during data transmission. In wireless personal area networks (WPANs) (IEEE 802.15.3c), codebook is designed in scenarios of indoor communications with the small number of antennas~\cite{magzine_solution}, where the beamforming protocol is similar to that in IEEE 802. 11ad. However, both of them only consider the analog beamforming (precoding).~\cite{scanning} proposed an hierarchical multi-resolution codebook based channel estimation for hybrid precoding/combining scheme. However, the proposed scheme may suffer from the destructive interference between the path gains when multiple paths are summed up in the earlier stages of the proposed algorithm \cite{scanning}.

\subsubsection{Proposed CS-Based Channel Estimation for MmWave Massive MIMO}
Some unique features of mmWave massive MIMO channels can be leveraged to alleviate the challenging problem of channel estimation.
\begin{itemize}
  \item
Due to the fixed BSs location, mmWave massive MIMO channels used for backhaul keep almost unchanged for a long time. Such long coherence time of channels indicates that channels are not necessary to be estimated very frequently compared with that used for RAN.
  \item
 The low-rank property of mmWave massive MIMO channel matrix indicates that although the dimension of mmWave massive MIMO channel matrix can be huge, its effective degrees of freedom (DoF) can be small. This inspires us to reconstruct channel matrix with significantly reduced measurements (sub-Nyquist sampling) under the framework of CS~\cite{Eldar}.

 \end{itemize}

By leveraging these features, we propose a CS-based channel estimation scheme as illustrated in Fig. \ref{fig:CE}, which consists of the following three phases:
\begin{itemize}
  \item
{\em Phase 1: coarse channel estimation}, as illustrated in Fig. \ref{fig:CE} (a), aims to acquire partial CSIT to generate the appropriate beamforming patterns for the following fine channel estimation with the improved received signal power. 
Specifically, the macro-cell BS sequentially broadcasts $L_{\rm{Ma}}$ predefined beamforming patterns in $L_{\rm{Ma}}$ successive time slots, 
while in every time slot,
each small-cell BS sequentially receives signal with $L_{\rm{Sm}}$ combining patterns in $L_{\rm{Sm}}$ successive sub-time slots. Then each small-cell BS feedbacks the indices of several optimal beamforming/combining patterns to the macro-cell BS.
  \item {\em Phase 2: channel estimation at small-cell BS,} as shown in Fig. \ref{fig:CE} (b), aims to estimate AoA and path gains at each small-cell BS. The macro-cell BS performs beamforming according to the feedback, while the $k$th small-cell BS estimates AoA and path gains by exploiting the finite rate of innovation (FRI) theory (analog CS)~\cite{Eldar}.
      With the aid of the predefined training signals ${\bf{S}}\in {\mathbb{C}}^{T_{\rm{gain}}^{\rm{Sm}} \times N_{\rm{BB}}^{\rm{Ma}}}$, the received signals at the small-cell BS is ${\bf{S}}{{\bf{P}}_d}{{\bf{P}}_a}{{\bf{A}}_T}{\bf{D}}{{\bf{B}}_R^*}{{\bf{C}}_a}{{\bf{C}}_d}$ according to (1), where the index $k$ is omitted, $T_{\rm{gain}}^{\rm{Sm}}$ is the time overhead. Since DPSN can disable some phase-shifters to set some elements of ${{\bf{C}}_a}$ to be zeros, the AoA and path gains estimation can be solved by the specific algorithms of FRI theory, e.g., estimating signal parameters viarotational invariance techniques (ESPRI) algorithm \cite{Eldar}.
  \item {\em Phase 3: channel estimation at macro-cell BS,} as shown in Fig. \ref{fig:CE} (c), aims to estimate AoD and path gains at the macro-cell BS. The specific procedure is similar to \emph{Phase 2}, where the $k$th small-cell BS transmits training signal while the macro-cell BS estimates channels.
\end{itemize}
The FRI theory  \cite{Eldar} can be used to accurately acquire the super-resolution estimation of AoA/AoD to effectively distinguish multiple paths with small angular difference, which can also relax the required resolution of the beamforming patterns in coarse channel estimation.
According to the estimated parameters and Eq.~(\ref{equ1}), both macro-cell BS and small-cell BSs can acquire the complete CSI for the following precoding/combining.  
\begin{figure*}[htb]
     \centering
     \vspace*{-4.0mm}
     \includegraphics[width=16.5cm]
     {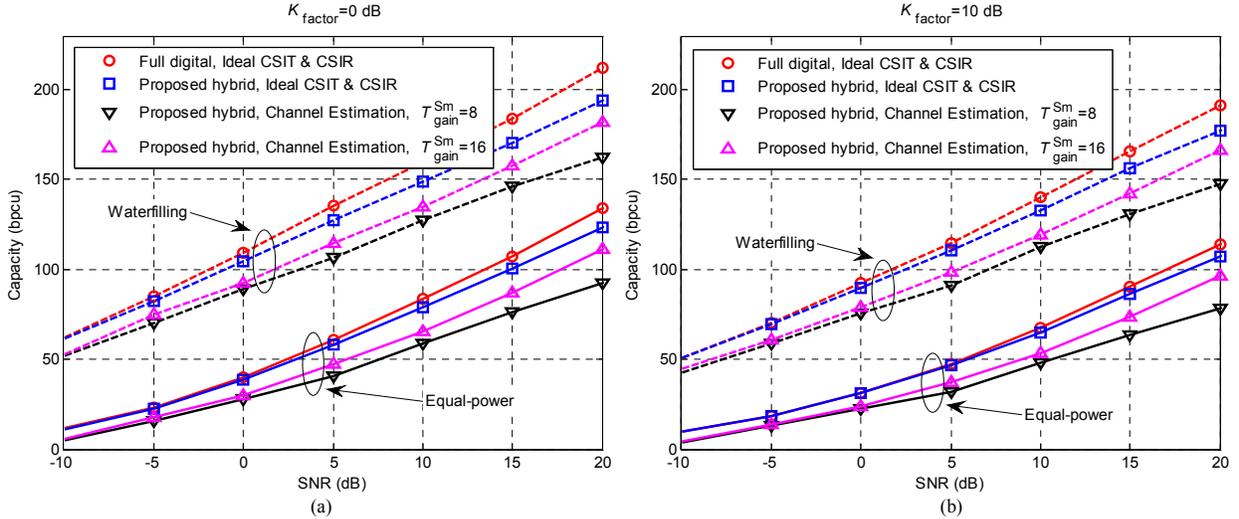}
     \vspace*{-4.0mm}
    \caption{Capacity comparison between the proposed hybrid precoding/combining scheme and the optimal full digital one: (a) $K_{\rm{factor}} = 0~{\rm dB}$; (b) $K_{\rm{factor}} = 10~{\rm dB}$.}     \label{sim}
    \vspace*{-4mm}
\end{figure*}

\subsubsection{Other Issues of Channel Estimation}
There still remains some problems to be investigated further, such as the optimal beamforming/combining patterns in coarse channel estimation \cite{scanning,magzine_solution}, training signals for AoA/AoD and path gains estimation \cite{Y_hybrid_training}, low-complexity high-accuracy CS-based channel estimation algorithms, effective channel feedback scheme, dynamic channel tracking to combat sudden blockage or slow channel changes. For instance, the microwave control link with only limited resource can be used to feedback the estimated parametric AoA/AoD,
since the number of AoA/AoD is typically 3$\sim$5 \cite{scanning}.
Regarding the CS-based channel estimation algorithm, in addition to FRI theory, other CS approaches such as low-rank matrix reconstruction are expected to be tailored for mmWave massive MIMO with low complexity \cite{Eldar}. Besides, the proposed channel estimation scheme (including the coarse channel estimation and the following estimation of AoA/AoD and path gains) is used to initially build the UDN backhaul link, where the latency can be negligible. Once the backhaul link is built, only the estimation of AoA/AoD and path gains is required to track the channels and then adjust the corresponding
precoding/combining, where the training sequences and data can be multiplexing in the time domain, so the latency can be also negligible.

    \vspace*{-3mm}
\subsection{Low-Complexity Hybrid Precoding/Combining for MmWave massive 3D MIMO}

For UDN in urban area, the precoding/combining scheme for mmWave massive 3D MIMO can exploit both azimuth and elevation to achieve the improved performance for backhaul link. Hence, ULA based hybrid precoding/combining scheme and the associated channel estimation proposed in this article should be extended to mmWave massive 3D MIMO in the future. Additionally, SVD and waterfilling may impose higher computational complexity on the hybrid precoding/combining in the 3D MIMO. Therefore, low-complexity hybrid precoding/combining schemes are desired for practical system design.
For instance, the spatial/angular sparsity of mmWave massive MIMO channels and the geometric structure of mmWave massive 3D MIMO may be exploited to reduce the complexity of SVD, while other sub-optimal power allocations can be considered to approach the performance of waterfilling with much low complexity.

   \vspace*{-3mm}
\subsection{Sampling with Low-Resolution ADC}
Small cells in future UDN can provide the Gbps user experience, which requires the large transmission bandwidth. To realize mmWave massive MIMO based backhaul, the cost of conventional high-speed ADC with high resolution can be unaffordable, while low-resolution ADC with low hardware cost is appealing. By far, 1-bit ADC based signal detector and precoding/combining have been investigated for mmWave massive MIMO \cite{magzine_solution,bit}. However, further efforts are still needed to generalize the associated results of 1-bit to more general cases, and constellation mapping, channel estimation, training signals, etc., may need to be reconsidered if low-resolution ADC is adopted.

   \vspace*{-3mm}
\section{Performance Comparison}\label{Performance_Comparison}
Fig. \ref{sim} compares the capacity (bit per channel use, bpcu) of the proposed DPSN based hybrid and the optimal full digital precoding/combining schemes in the downlink, where both the waterfilling power allocation and equal-power allocation are investigated.
In simulations, ULA is considered at both macro and small-cell BSs, the working frequency is $60~\rm{GHz}$, $K=4$, $N_{\rm a}^{\rm{Ma}}=512$, and $N_{\rm a}^{\rm{Sm}}=32$.
For the optimal full digital scheme, $N_{\rm{BB}}^{\rm{Sm}}=N_{\rm a}^{\rm{Sm}}$ and $N_{\rm{BB}}^{\rm{Ma}}=N_{\rm a}^{\rm{Ma}}$, where the ideal CSIT and CISR are assumed as the upper bound of capacity. In the proposed scheme, $N_{\rm{BB}}^{\rm{Sm}}=4$ and $N_{\rm{BB}}^{\rm{Ma}}=16$, where cases of ideal CSI known by transceiver and unideal CSI acquired by the proposed CS-based channel estimation scheme are considered. 
For mmWave massive MIMO channels, $L$ in simulations follows the discrete uniform distribution ${\cal{U}}_d[2,6]$, and AoA/AoD follow the continuous uniform distribution ${\cal{U}}_c[0,2\pi)$. For path gains, we consider Rican fading consisting of one LOS path and $L-1$ equal-power NLOS paths, where path gains follow the mutually independent complex Gaussian distribution with zero means, and $K_{\rm{factor}}$ denotes the ratio between the power of LOS path and the power of NLOS path.

Fig. \ref{sim} shows that the proposed hybrid scheme with ideal CSIT and CSIR suffers from a negligible capacity loss
compared with the optimal full digital scheme, although the proposed scheme only uses a much smaller number of BB chains.
This is because the proposed scheme exploits the low-rank property of mmWave massive MIMO channel matrix, where capacity
exhibits ceiling effect when the number of BB chains are sufficiently large. Moreover, with the increased $T_{\rm{gain}}^{\rm{Sm}}$,
capacity of the proposed scheme with CS-based channel estimation approaches that with the ideal CSI. This is because the increased
number of measurements can improve the channel estimation performance. Besides, schemes with waterfilling power allocation outperform
these with equal-power allocation, which indicates that waterfilling or other power allocations should be considered in practical
system design for the improved backhaul capacity.

    \vspace*{-2mm}
\section{Benefits and Opportunities of MmWave Massive MIMO Based Backhaul Network}\label{benefits}

\subsection{Point-to-Multiple-Points Backhaul}
In conventional wireless backhaul network, point-to-point (P2P) and P2MP are two typical network topologies. A general consensus is P2MP has the lower total cost of ownership than that in P2P \cite{in_band}. By far, P2P has been widely used in both microwave and mmWave backhaul systems, while P2MP has been implemented in sub 6 GHz licensed band. However, there are no satisfactory P2MP based backhaul solutions in mmWave, which is urgently desired by industry. In this article, the proposed mmWave massive MIMO based wireless backhaul enables a macro-cell BS to simultaneously support multiple small-cell BSs, which can provide the viable approach to realize the P2MP backhaul in mmWave.
    \vspace*{-2mm}
\subsection{Beam Division Multiplex (BDM) Based Scheduling}
Time division multiplex (TDM) based scheduling has been proposed for mmWave based backhaul \cite{in_band}. However, this scheme may suffer from the latency, since backhaul links between different small-cell BSs and macro-cell BS are multiplexing in different time slots. In this article, the proposed mmWave massive MIMO based backhaul network can realize the BDM based scheduling due to the flexible spatial multiplexing and hybrid beamforming. In the practical backhaul network, according to the backhaul load, the macro-cell BS can flexibly combine the TDM and BDM to support more small-cell BSs. For instance, links with heavy load or without LOS path can be assigned more beam resources, while multiple links with light load can be multiplexing in the time domain. Additionally, TDM based in-band mmWave backhaul is recently proposed for the reduced cost, where backhaul and RAN share the same frequency band \cite{in_band}. Obviously, with the proposed mmWave massive MIMO scheme, the BDM based scheduling may be another competitive solution for the in-band mmWave backhaul with lower latency.
    \vspace*{-2mm}
\subsection{TDD is Suitable for MmWave Backhaul Network}
For frequency division duplex (FDD)-based mmWave backhaul, the uplink and downlink have to use different bands. However, the regulation in mmWave may be different in different countries. This indicates that one single device may be not suitable in various countries. By contrast, the uplink and downlink in TDD share the same band. Hence one single device can be employed in various countries. Moreover, since different operators will employ UDN in the same areas, the mutual interference of backhaul networks must be considered. Compared to FDD with the different uplink/downlink channels, TDD is more easier to find clean spectrum and avoid interference. Moreover,
since the asymmetric traffic is dominant in backhaul network, TDD can flexibly adjust the ratio of
time slots in the uplink and downlink according to the traffic requirement \cite{TDD}. For the practical TDD mmWave massive MIMO based backhaul, the adaptive interference management is desired to avoid mutual interference of different operators' UDN, and automated configuration solutions are expected for the plug-and-play backhaul network, especially for unlicensed V-band. While for licensed E-band, spectrum regulation needs to be further improved.

    \vspace*{-3mm}
\section{Conclusions}
This article discusses a promising wireless backhaul based on mmWave massive MIMO for future 5G UDN.
We have explored the fundamental issues of the implementation of the mmWave massive MIMO for wireless backhaul. 
Especially, by leveraging the low-rank property of mmWave massive MIMO channel matrix, we propose the DPSN based hybrid precoding/combining scheme and the associated CS-based channel estimation scheme. The proposed scheme can guarantee the macro-cell BS to simultaneously support multiple small-cell BSs with multiple streams for each small-cell BS. This is essentially different from conventional hybrid precoding/combining used for RAN.
The proposed scheme can provide the viable approach to realize the desired P2MP backhaul topology and novel BDM based scheduling, and it may also facilitate the in-band backhual in mmWave.
Additionally, some potential research directions to enable mmWave massive MIMO based wireless backhaul are highlighted, which may become active research topics in near future.

\end{document}